# Photoelectric response of localized states in silica glass

A.N. Trukhin[1]

Institute of Solid State Physics, University of Latvia, Latvia

Abstract

The photoelectric response of pure silica glasses excited by excimer lasers has been studied. The samples were made under various conditions. Photoelectric polarization of samples due to the Dember effect has been registered. The signal was recorded under the conditions of a space charge limited current. The space charge resulting from the capture of electrons and holes creates a static electric field that prevents diffusion of released charge carriers. The current registration in the external circuit stops, despite the continuation of photoexcitation. This effect was used as a fact of measuring the photocurrent in the sample volume instead of the parasitic current that is not associated with the sample. The screen has been chosen to prevent the influence of a spurious signal. It has been found that charge carriers are released when excited in the spectral absorption range of localized states of silica. Based on the Dember effect, the sign of the photoelectric response shows the type of charge carriers - an electron or a hole is mobile. Thus, a sample containing aluminum without alkali ions gives a negative signal, which indicates the diffusion of electrons at 290 K, since aluminum is an effective hole trap. An oxygen-deficient sample at 290 K provides a positive signal indicating the diffusion of holes, because the center of oxygen deficiency is an effective electron trap. This sample at 100 K provides a negative signal due to the effective self-trapping of holes.

INTRODUCTION

The localized states of silica glass have been proved only in recent decades [1, 2], but their existence is predicted by the general properties of disordered materials. Previously the main objects of research were heavy doped silica glass with additional silicon or aluminum. For these samples, it was found that when the localized states of silicon dioxide are excited, the holes appear as self-trapped, and the electrons are trapped in the so-called oxygen deficient center (I) or ODC(I) [3]. Earlier it was noted that the concentration of self-trapped hole centers increases with increasing oxygen deficiency [4]. Recombination of a hole with an electron on the ODC (I) provides luminescence of the modified oxygen deficient center. The modified oxygen-deficient center is not a point defect. Point defect related to oxygen deficiency is twofold coordinated silicon centers or ODC(II), which in itself never participates in the recombination of the electron-hole system [2, 3]. ODC (II) is excited only in the intracenter process. In fact, the structure of modified oxygen deficient defects is not yet understood. This should be a complex structure, including two-fold coordinated silicon appearing as a transient in process of electron-hole recombination. In the case where the silica contains technologic impurities such as chlorine, fluorine or an OH group, the oxygen deficient center modified by these impurities has properties similar to those observed without the mentioned impurities.

---

[1] Corresponding author.
E-mail address: truhins@cfi.lu.lv

Thus, the excitation of localized states predicts some electrical activity associated with the possible release of charge carriers. Then, it is necessary to study the photoelectric response of the insulator.

When the insulator is illuminated with light, two electrical effects can be observed - external photoelectron emission and photoconductivity. The first involves the diffusion of the released charge carriers to the volume of the insulator, and then their escape from the insulator to the vacuum (the bulk external photoelectric effect), [5]. The bulk external photoelectric effect is determined by the properties of the energy bands of the insulator - the density of electronic states in the valence and conduction bands, so it can be activated in the range of intrinsic absorption of the material. Photoconductivity is associated with the release of charge carriers and their movement in the volume of materials. In the case of an insulator, the diffusion of carriers through the volume can take place for extremely thin samples. Typically, the current in the external circuit is provided by shifting the carriers a distance to recombination or capture. Gudden and Pohl [6] (see, for example in [7]) were pioneers in the study of photoconductivity in insulators. They introduced the concept of primary and secondary currents. The primary current is connected to the charge carriers released by the light inside the sample, but the secondary current is a reaction to the primary current. Thus, the contact can be between the insulator and the electrodes on it [8] after releasing the charge carriers. Then the injection of charge carriers from the electrodes can lead to a strong increase the concentration carriers in the insulator. Injected charge carriers make the picture of the observed phenomena more complicated. The main processes that determine the internal photoelectric effect are the diffusion of charge carriers and their trapping in traps. In the case of a strong difference in the values of the diffusion coefficient of electrons relative to the hole, their separation occurs in space, and as a result, an electric field appears in the bulk of the insulator (Dember field, [9]). The trapping of carriers on traps induces a space charge. The space charge field prevents the carriers from moving. The Dember field can produce a current in the external circuit, because its nature is due to dynamic causes. On the contrary, the space-charge field is static (charge carriers are localized) and, thus, it stops the current in the external circuit, and the Dember field looks like frozen. This phenomenon was called photoelectric polarization [10] and for the study of the frozen Dember field a vibration electrode was proposed. Previously, the photoelectric response of silicon dioxide in the optical transparency range was studied in [11] for samples with heavy doping of excess silicon. Thus, in both cases, external electron emission and the release of a charge in a volume, the space charge can stop the current in the external circuit. Then, on the light pulse of the Π-form, we obtain a damped photoelectric response during the light pulse. Even in some cases, the next light pulse does not provide a pulse of the photoelectric response. This depends on the lifetime of the space charge. The latter depends on the recombination of charge carriers.

Now experiments are conducted with pure silica samples without heavy doping with additional silicon. The main attention was paid to samples with a small optical absorption below the optical gap (~ 8 eV). These samples are made intentionally in excess conditions of oxygen and / or fluorinated. To understand the role of OH groups, a sample of pure silica of type III was also studied. To isolate the role of the electron and hole in the photoelectric response, a silica sample doped with aluminum without alkali ions was measured.

EXPERIMENTAL

The samples of the study were pure silica glasses of type III (trademarks KY-1, Corning 7940) and type IV (trademarks KC-4B or KS-4V with different levels of oxygen deficiency, including excess oxygen samples and intentionally activated with chlorine, fluorine or aluminum). Type III synthetic silica is made from SiCl4 combusted in an oxygen-hydrogen flame. KS-4V type IV silica was prepared by electrofusion of crystallized silicon dioxide [12]. Crystallization was done using lithium as a catalyst. Lithium was washed with chlorine, which was removed with oxygen. Melting before oxygen treatment gave chlorine-containing samples. F-doping was carried out by melting in an atmosphere of pure $SiF_4$. The concentration of fluorine in the doped sample is 0.1 wt.%. Before electrofusion, another sample of quartz glass was doped with 0.01 - 0.05 wt.% $Al_2O_3$. Figure 1 shows the measurement scheme. The sample was glued with a silver dag on the holder. On the other surface of the sample a contact was made with a silver dag. As an electrometer, a dynamic capacitor electrometer VA-J-52, made in the 60's in the RTF of East Germany, was used. The data of the electrometer were read by an Agilent voltmeter connected to a PC. We used a liquid nitrogen cryostat with the possibility of connecting an electrometer.

Excimer lasers (KrF-5 eV or 248 nm, ArF – 6.4 eV or 193 nm and $F_2$ – 7.9 eV or 157 nm), model PSX-100, were made by Neweks, Estonia has a pulse energy of about 5 mJ ($F_2$ laser 0.5 mJ) with a duration of 5 ns used for excitation.

Experimental results are presented as obtained and the noise reflects the measurement accuracy.

RESULTS

The optical absorption spectra of the samples are shown in Fig. 2, 3. Samples containing excess oxygen and fluorine have a minimum absorption level below the optical gap at 8.2 eV. A sample with chlorine has a pronounced band at 7.6 eV, Fig.2. Samples with OH groups have a small absorption level below 7.5 eV and above this energy the samples are not transparent, Fig.2. Samples with aluminum, Fig. 3, have an increase in absorption above 7 eV. This absorption is proportional to the level of added $Al_2O_3$ before the fusion of the sample. Fig. 4 shows the photoelectric response pulse of a sample of silica glass with excess oxygen and excitation by an excimer laser $F_2$ (157 nm) at 290 K. It is found that under the condition of an experiment with excitation for 400 s, the current in the external circuit slowly increases and then decreases to zero, despite the continuation of excitation. The next excitation pulse provides a very small current. Full recovery is possible after several days of waiting. Thus, it can be established that the charge is released with some delay since the start of the excitation. The movement of the charge causes a current in the measuring circuit. The decrease in time is due to the restriction of diffusion due to space charge. This decrease indicates a released charge trapping and the creation of a spatial field that impedes the movement of the newly released charge. By the sign of the photoelectric response, it was possible to establish a higher mobility of the hole, respectively, of the Dember effect. A positive charged hole diffuses for a greater distance than the electron, and then provides a positive response in the measuring circuit in Fig.1.

An experiment on the effect of temperature on the photoelectric response is presented in Fig.5. First, it should be emphasized that the sample remembers the previous excitation pulse even after several hours of waiting. Therefore, the current level decreases, but the shape of the pulse with increasing and decreasing with continuous excitation remains at 290 K. Cooling the sample to low temperatures causes an electrical signal without excitation. This can be explained by the changes in the cryostat with the cooling causing the effect of the microphone. While waiting for a sufficiently long time to reduce this parasitic electrical response, at the same time, the effect of space charge after the previous excitation pulse decreases. The photoelectric response obtained at low temperature is initially negative. This can be explained by self-trapping holes at such temperatures (see, for example, [4, 13]). The subsequent positive response can be explained by some equilibrium of self-trapping of holes and electron capture. In Fig. 6 shows the photoelectric response pulse at 290 K for aluminum doped silicon dioxide. Strong negative pulses are observed for $F_2$ and ArF lasers. Both pulses show the limitation of the current by space charge. It is known that aluminum in silica glass is an effective trap for holes [13] and, accordingly, the Dember effect shows more efficient diffusion of electrons. Naturally, higher photon energy provides a higher signal.

A complex photoelectric response was obtained for fluorine doped silicon dioxide. Despite the fact that low values of the absorption coefficient excitation gives pulses of the photoelectric response, Fig.7. An initial positive pulse was obtained for the $F_2$ laser. Continuous excitation indicates a change in the sign of the response. Then the response remains high, increasing during excitation. Only negative pulses were obtained for the ArF laser. Laser excitation of KrF provides a low-intensity photoelectric response independent of time, and then the absence of charge capture to create a space charge limited current. Then, probably, when the KrF laser is excited, there is no charge release in the sample, and only some parasitic signal is observed.

A complex picture, when the $F_2$ laser excitation manifests itself in the complication of the charge-release and capture process with the change in the type of mobile carriers. Also, samples with chlorine exhibit complex processes during excitation, Fig.8. Initially, a negative photovoltaic response pulse was obtained. Obvious is the creation of a space charge limiting current during prolonged excitation. The predominance of electronic mobility was observed. Cooling to 150 K (slightly lower than the self-trapping of the holes) the response pulse is positive with a strong effect of limited current with space charge. Then the electrons are effectively captured in the samples with chlorine. Subsequent measurement at 290 K provides a positive pulse also with the effect of the space charge limiting current. Thus, when the temperature is changed in a sample, complex reconstructions occur with some memory effect.

Silicon dioxide type III (OH) also exhibits a photoelectric response, Fig. 9. For 7.9 eV (157 nm) $F_2$ laser, the absorption coefficient is high, and the sample under such conditions very quickly creates a space charge limiting current. The space charge saturates after about 10 s. The signal is positive, hence the electrons are captured very effectively. The ArF laser (6.4 eV or 193 nm), whose absorption coefficient is relatively small, also provides a positive pulse with less space charge creation effect. The KrF-laser (5 eV or 248 nm), as in the previous cases, does not excite the charge release in the sample, basically it provides a certain parasitic signal.

DISCUSSION

The photoelectric response of samples of pure silica glass was studied using three excimer lasers to excite photoelectric response below the optical gap. The response includes physical data and a parasitic signal. It is assumed that the parasitic signal does not change with time under conditions of constant geometry. The possible parasitic signal is explained by two factors. One is the scattered light incident on the sensitive electrode of the electrometer. In Fig. 7 the case of KrF laser excitation could be estimated as only parasitic signal. Some light scattered on the screen could fall on sensitive electrode. The light incident on the sensitive electrode gives a positive signal sign. Another parasitic signal is the photoelectron emission from the screen also due to the transmission of light through the sample. The signal in this case also does not change in time, but the sign of the signal is negative. In the same Fig. 7, a negative signal, slightly varying in time when ArF is excited, may be associated with photoemission from the screen wall.

A signal with a physical meaning is taken into account if it changes with time. It grows at the beginning of excitation, and then falls in any case to zero, despite continuous excitation (e.g. Fig.4).

Interpretation of this signal consists in the release of charge carriers due to the absorption of light in the sample. The diffusion of carriers provides a current in the external circuit. The drop in the response is due to the trapping of charge carriers, which static field limits the current. This space charge is created by electrons and holes captured by some defects. The initial increase in the intensity of the signal, Fig. 4, may indicate competition between two processes: the release of a charge with diffusion of charge carriers and charge capture with the creation of a space charge that limits diffusion of carriers. Initially, the charge volume is small, and the response increases with time. Then the resulting space charge limits diffusion, and growth slows down. The decline began when the level of the space charge became large enough to slow the diffusion. When diffusion is stopped by the space charge, the current in the external circuit becomes zero, despite continuous excitation.

The sign of the photoelectric response is defined as charge carriers with a higher diffusion capacity. It depends on the nature of the traps. Aluminum in silica effectively absorbs holes, and then the sign of the signal is negative because of the prevailing diffusion of the electron. Once the sign of the reaction changes during excitation and that the means of photo-stimulated reconstruction and the creation of traps, as was seen in the case of a sample doped with fluorine, Fig.7.

CONCLUSIONS

It is found that, despite the low absorption coefficient of samples of pure silica glass, excitation by means of excimer lasers ensures the release of charge carriers. Electron and holes are released, diffused and trapped in all studied samples. Trapping leads to space charge creation limiting mobility of liberated electron and holes. Therefore, excited localized states of silica provide mobile charge carriers upon excitation.

ACKNOWLEDGMENTS

This work was supported by the Latvian Science Council Grant No lzp-2018/1-0289.

References

[1] A.N.Trukhin, Excitons, localized states in silicon dioxide and related crystals and glasses, International school of solid state physics, 17th course., NATO science series. II Mathematics, Physics and Chemistry  Defects in SiO2 and related dielectrics: science and technology, Ed D.Griscom, G.Pacchioni, L.Skuja, Kluwer Academic Publishers, Printed in the Netherlands, 2 (2000) 235-283.

[2] A.N. Trukhin, Luminescence of localized states in silicon dioxide glass. A short review, Journal of Non-Crystalline Solids 357 (2011) 1931–1940.

[3] A. N. Trukhin, J. Teteris, A. Fedotov, D.L.Griscom, G.Buscarino, Photosensitivity of SiO2-Al and SiO2-Na glasses under ArF (193 nm) laser, Journal of Non-Crystalline Solids, 355 (2009) 1066-1074.

[4] Trukhin, A.N.  Troks, J. Griscom, D.L.  Thermostimulated luminescence and electron spin resonance in X-ray-and photon-irradiated oxygen-decient silica, J. Non-Cryst. Solids. 353 (2007) 1560-1566.

[5] Spicer W. E. Photoemission and band structure. In: Survey of phenomena in ionized glasses. Vienna, 1968, p. 271-289.

[6] B.Gudden und R.Pohl, Uber lichtelektrische Leitffafigkeit von Diamanten. Zeit. f. Phys., 1920, 3, 123-124.

[7] N. Mott, R. Gurney, Electronic processes in Ionic Crystals, Oxford, 1958.

[8] M. Lampert and P. Mark, Current injection in solids, Academic press N.-Y and London, 1970

[9] H. Dember. "Über eine photoelektronische Kraft in Kupferoxydul-Kristallen (Photoelectric E.M.F. in Cuprous-Oxide Crystals)". Phys. Zeits. 32 (1931) 554.

[10] I. Vitol, I. Tale, Investigation of photoelectric polarization of dielectric by dynamic capacitor method, Institute of Physics and Astronomy of the Academy of Science of Estonian SSR, 1959, Nr10, p.220-238.

[11] A.N. Trukhin, H.-J. Fitting, Investigation of optical and radiation properties of oxygen deficient silica glasses, J. Non-Cryst. Solids,, 248 (1999) 49-64.

[12] A.G. Boganov, E.M. Dianov, L.S. Kornienko, E.P. Nikitin, V.S. Rudenko, A.O. Rybaltovsky, P.V. Chernov, Water-free silica glass for optical fibers and radiation and optical properties,  Quantum Electronics (Sov.) 4 (1977) 996-1003.

[13] D.L.Griscom, The nature of point defects in amorphous silicon dioxide, 17th course., NATO science series. II Mathematics, Physics and Chemistry  Defects in SiO2 and related

dielectrics: science and technology, Ed D.Griscom, G.Pacchioni, L.Skuja, Kluwer Academic Publishers, Printed in the Netherlands, 2 (2000) 73-159.

Figures caption

Fig.1

Photoelectric response measurement circuit. 1 – sample, 2 – silver dag, 3 – electrometer, 4 – copper sample holder, 5 – Teflon collimator preventing electron emission from metallic part of the system, 6- excitation light, 7 - copper screen, 8 –ground.

Fig.2

Optical absorption of pure silica glasses type III and type IV. F- fluorine doped, 161B – oxidized and 155B chlorine containing as well as wet silica type III with OH groups.

Fig.3

Optical absorption spectra of studied aluminum doped silica glass with 0.01 - 0.05 wt.% $Al_2O_3$ added before fusion. Also spectral positions of ArF and $F_2$ excimer laser photons are shown as well.

Fig.4

Photoelectric polarization of oxygen doped silica under pulses of F2 laser. Sample without previous irradiation.

Fig.5

Photoelectric polarization of oxygen doped silica under pulses of $F_2$ laser at different temperatures.

Fig.6

Photoelectric polarization of Al doped silica at 293 K.

Fig.7

Photoelectric polarization of fluorine doped silica.

Fig.8

Photoelectric polarization of chlorine doped silica under pulses of F2 laser at different temperatures.

Fig.9

Photoelectric polarization of OH doped silica type III (KY-1) under three excimer lasers.

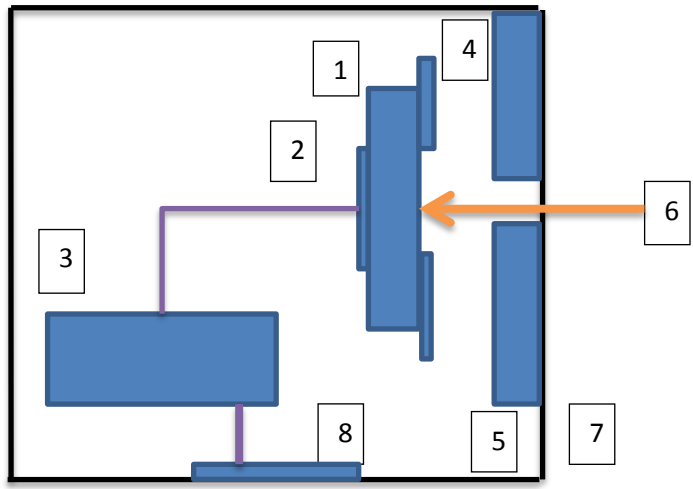

Fig.1

Photoelectric response measurement circuit. 1 – sample, 2 – silver dag, 3 – electrometer, 4 – copper sample holder, 5 – Teflon collimator preventing electron emission from metallic part of the system, 6- excitation light, 7 - copper screen, 8 –ground.

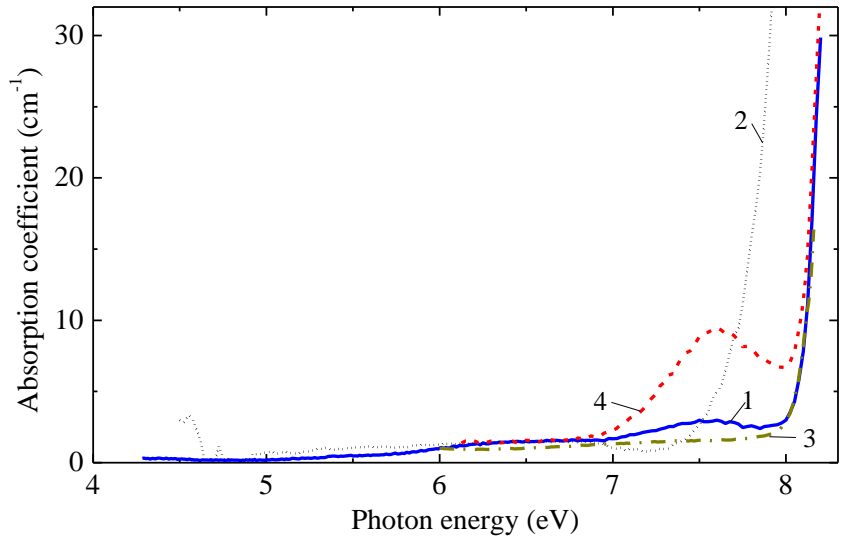

Fig.2

Optical absorption of pure silica glasses type III and type IV. 1 – excess $O_2$, 2 – OH containing type III Corining 7940, 3 - fluorine doped, 4 - $Cl_2$ containing.

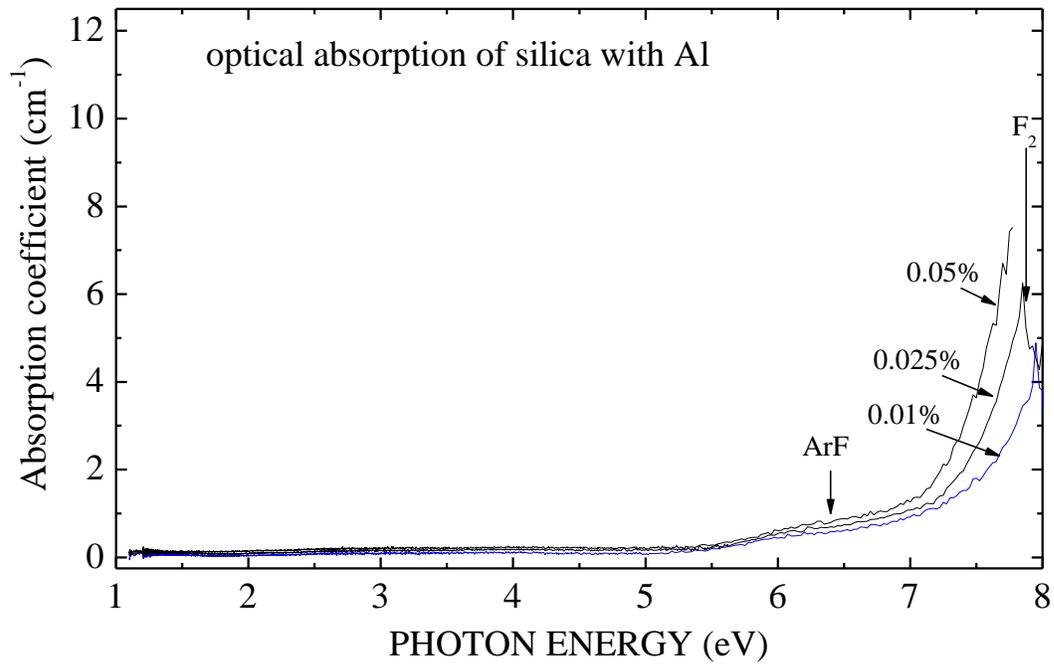

Fig.3

Optical absorption spectra of studied aluminum doped silica glass with 0.01 - 0.05 wt.% $Al_2O_3$ added before fusion. Also spectral positions of ArF and $F_2$ excimer laser photons are shown as well.

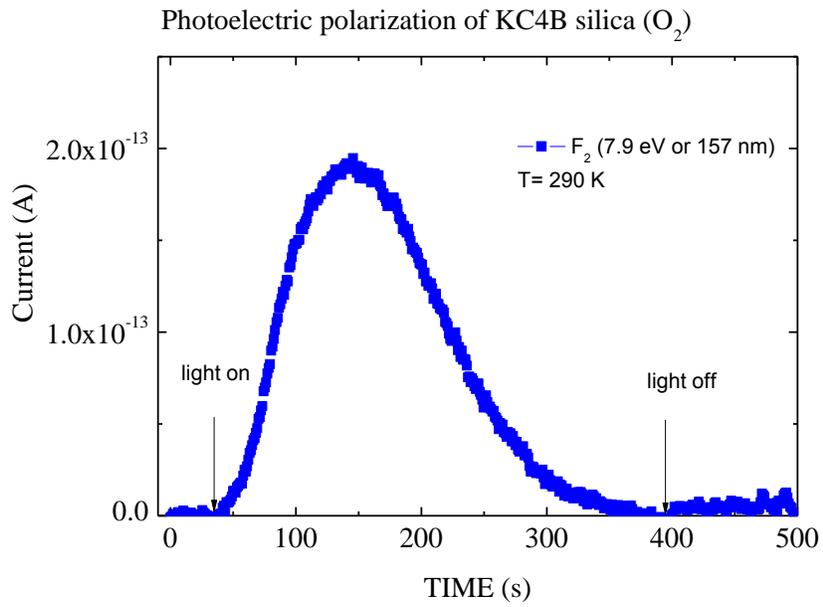

Fig.4

Photoelectric polarization of oxygen surplus silica under pulses of F2 laser. Sample was not previously irradiated.

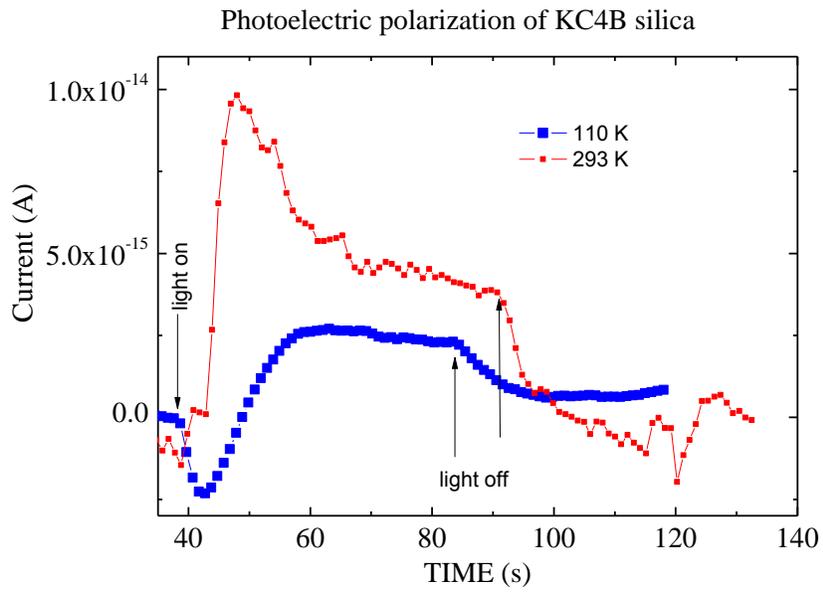

Fig.5

Photoelectric polarization of oxygen surplus silica under pulses of $F_2$ laser at different temperatures.

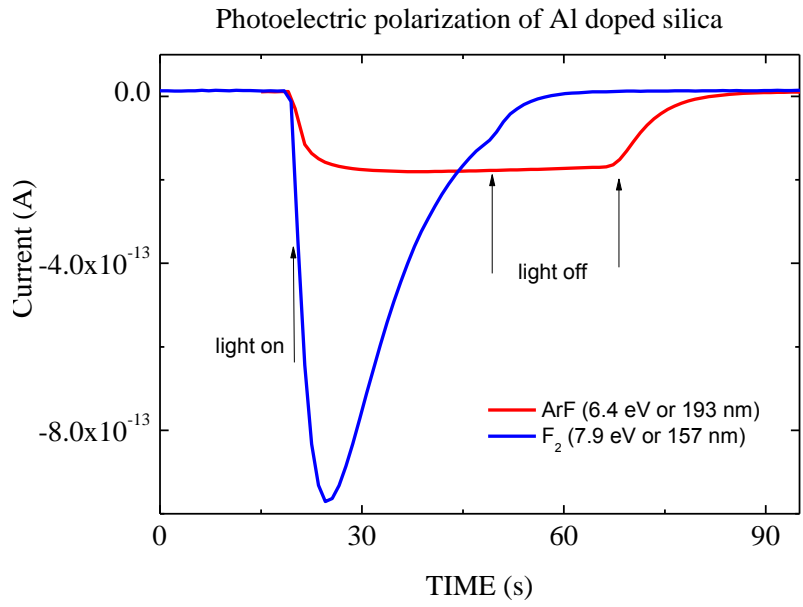

Fig.6

Photoelectric polarization of Al doped silica at 293 K.

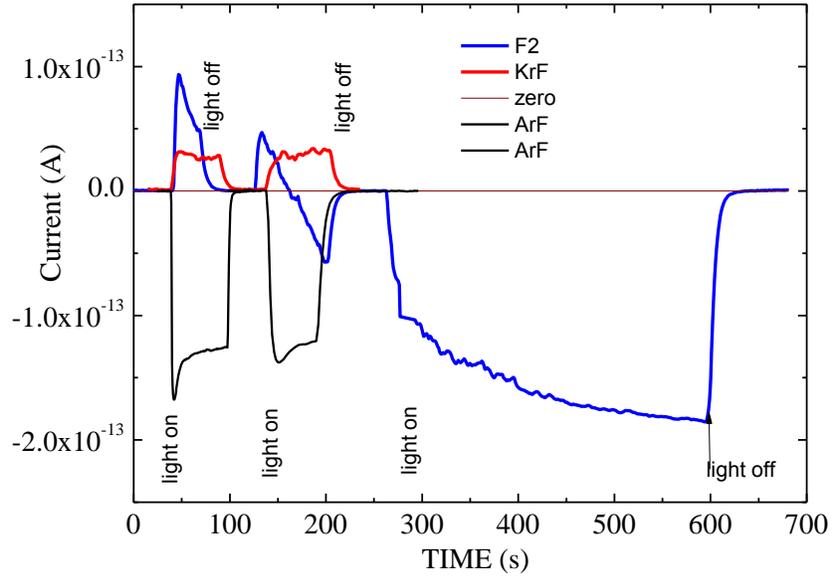

Fig.7

Photoelectric polarization of fluorine doped silica.

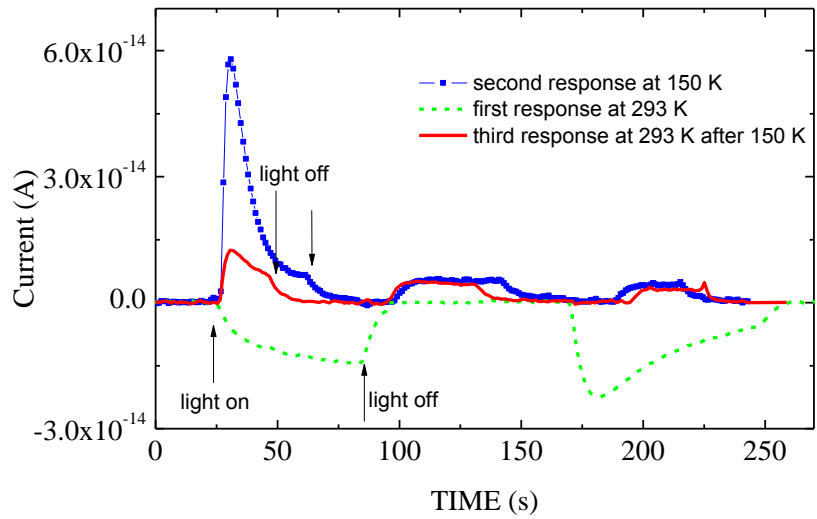

Fig.8

Photoelectric polarization of chlorine doped silica under pulses of $F_2$ laser at different temperatures.

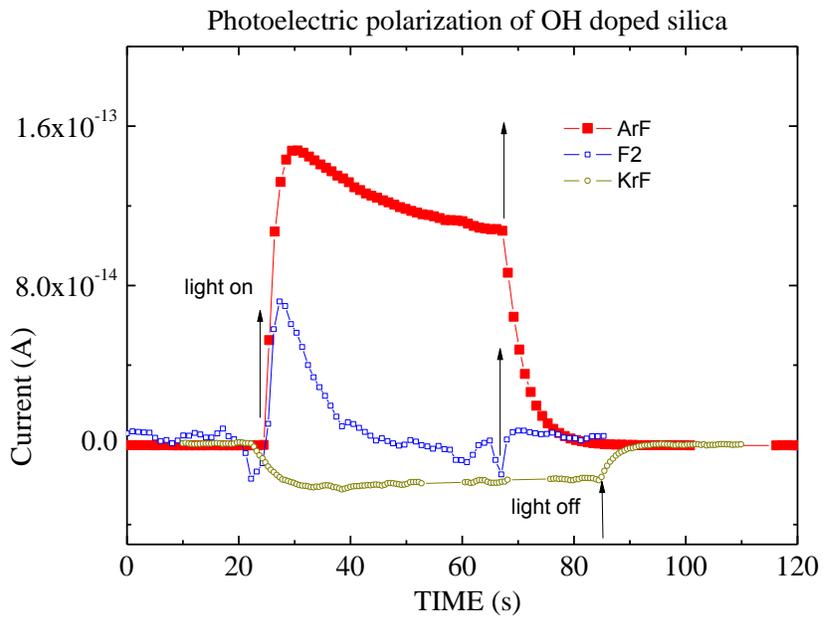

Fig.9

Photoelectric polarization of OH doped silica type III under three excimer lasers.